\documentstyle[psfig]{mn}

\title{{ An Artificial Neural Network Approach to Classification of Galaxy Spectra}}
\author[S.R. Folkes, O. Lahav and S.J. Maddox]
       {S.~R.~Folkes,$^1$ O.~Lahav$^1$ and S.~J.~Maddox$^2$ \\
$^1$ Institute of Astronomy, The Observatories, Madingley Road, Cambridge, CB3 0HA \\
$^2$ Royal Greenwich Observatory, Madingley Road, Cambridge, CB3 0EZ}
\date{Accepted 
      Received}
\pubyear{1996}

\begin{document}

\maketitle

\begin{abstract}
We present a method for automated classification of galaxies with low
signal-to-noise (S/N) spectra typical of redshift surveys. We develop
spectral simulations based on the parameters for the 2dF Galaxy
Redshift Survey, and with these simulations we investigate the
technique of Principal Component Analysis when applied specifically to
spectra of low S/N. We relate the objective principal components to
features in the spectra and use a small number of components to
successfully reconstruct the underlying signal from the low quality
spectra. Using the principal components as input, we train an
Artificial Neural Network to classify the noisy simulated spectra into
morphological classes, revealing the success of the classification
against the observed $b_{\rm J}$ magnitude of the source, which we
compare with alternative methods of classification.  We find that more
than 90\% of our sample of normal galaxies are correctly classified
into one of five broad morphological classes for simulations at
$b_{\rm J}$=19.7. By dividing the data into separate sets we show that
a classification onto the Hubble sequence is only relevant for normal
galaxies and that spectra with unusual features should be incorporated
into a classification scheme based predominantly on their spectral
signatures. We discuss how an Artificial Neural Network can be used to
distinguish normal and unusual galaxy spectra, and discuss the
possible application of these results to spectra from galaxy redshift
surveys.

\end{abstract}


\section{Introduction}
In the near future new galaxy surveys will provide a large number of
spectra, which will enable important measurements of galaxy
properties. For example, the 2 degree field (hereafter 2dF) Galaxy
Survey aims to collect 250,000 spectra. The integrated spectrum of a
galaxy is a measure of its stellar composition and gas content, as
well as its dynamical properties. Moreover, spectral properties often
correlate fairly closely with galaxy morphology. Indeed, as the
spectra are more directly related to the underlying astrophysics, they
could prove a more robust classifier for evolutionary and
environmental studies.  Spectra can be obtained to larger redshifts
than ground-based morphologies and, as 1-D data sets, are easier to
analyse. Although the concept of spectral classification goes back to
Humason (1936) and Morgan \& Mayall (1957), few uniform data sets are
available and a number of different approaches to the problem are
possible.

Spectral classification is important for several practical and
fundamental reasons.  In order to derive luminosities corrected for
the effects of redshift, the $k$-correction must be estimated for each
galaxy. The rest-frame spectral energy distribution is needed, which
can be obtained by matching the observed spectrum against templates of
local galaxies.  The proportion of sources in each class as a function
of luminosity and redshift is of major interest. Apart from its
relevance for environmental and evolutionary studies, new classes of
objects may be discovered as outliers in spectral parameter space.
Furthermore, by incorporating spectral features with other parameters
(e.g. colour and velocity dispersion) an `H-R diagram for galaxies'
can be examined with possible important implications for theories of
galaxy formation.

In this paper we explore the PCA and Artificial Neural Network
(hereafter ANN) combination when applied to noisy galaxy spectra. PCA
has been demonstrated to be a useful tool for spectral classification,
with applications to stellar spectra (e.g. Murtagh \& Heck 1987; von
Hippel et al. 1994), QSO spectra (Francis et al. 1992) and galaxy
spectra (Sodr\'e \& Cuevas 1994; Sodr\'e \& Cuevas 1996; Connolly et
al. 1995). ANNs have been used for classification of images
(Storrie-Lombardi et al. 1992; Naim et al. 1995; Lahav et al. 1995)
and stellar spectra (Storrie-Lombardi et al. 1994) along with a
variety of other astronomical applications. Other approaches have also
been taken, such as analysing the weight of specific components in
each galaxy spectrum (Zaritsky et al. 1995). This approach is similar
to the PCA technique, but the templates do not form an orthogonal set,
although they can be chosen specifically to highlight certain
characteristics in the spectra, such as young stars or emission
lines. However, this approach does not allow for spectral variations
extending outside the scope of the predetermined template set.

In this paper we use PCA on the complete spectra of the data set as
opposed to some other spectral analyses which use specific measured
quantities from the spectra (e.g. line strengths). We prefer to use
the complete data so that we are not restricted to a set of
predetermined measures. By using all the available data, the S/N
inherent in the method is increased.

ANNs, originally suggested as simplified models of the human brain,
are computer algorithms which provide a convenient general-purpose
framework for classification (e.g. Hertz et al. 1991).  ANNs are
related to other statistical methods common in Astronomy and other
fields.  In particular ANNs generalize Bayesian methods,
multi-parameter fitting, PCA, Wiener filtering and regularisation
methods (e.g. Lahav et al. 1996).

We take the approach of using a fairly small set of high S/N spectra,
and degrade them using the parameters of the 2dF system on the AAT.
This produces simulated spectra for a range of possible noise levels
which allows us to quantify the affect of the increasing noise and put
limits on the success rates we hope to achieve for the spectral
classification. In section 2 we describe the data set and show
examples of the simulations. In section 3 we utilize the technique of
Principal Component Analysis to compress the data set and to extract
the `real' information from the noisy spectra, leading to section 4,
where we look at the spectral reconstructions based on the PCA,
highlighting the ideal methods to use. In section 5 we use an
Artificial Neural Network to operate on the results from the PCA, and
demonstrate the level of classification success attained by this
method. We end with a discussion of the results and the conclusions of
the investigation.

\section[]{Data}
The spectra used in this investigation are taken from the
spectrophotometric atlas of galaxies (Kennicutt 1992) and represent
the integrated spectra of local galaxies. They have been selected to
demonstrate a wide range of spectral signatures. Most of the spectra
have 5-8\AA \ resolution but a few have been observed giving a lower
resolution of 10-25\AA . The spectra cover the wavelength range from
3650-7100\AA , although for the purposes of this paper, we are left
with a slightly shorter range when the simulation process has been
applied (see Appendix A). More details of the observations are given
in Kennicutt (1992). For the purposes of this paper the spectra have
been split into two groups. The `Normal26' spectra have been
selected as being representative spectra for galaxies of normal
morphological type, i.e.  galaxies which conform simply to the Hubble
classification scheme (Hubble 1936). The `Unusual29' spectra
comprise the remainder of the galaxies observed by Kennicutt. These
spectra include peculiar and starburst galaxies and also galaxies with
Seyfert nuclei.
\begin{table}
\caption{Morphological groups. Each group is given a name, a number
G (for use in section 5), the T-Types covered by the group and the
percentage of the galaxies in the ESO catalogue which fall into that
group.}
\begin{tabular}{llcl}
\hline
{\em Group} & {\em G} & {\em T-Types} & {\em ESO \%}\\ \hline
\hline
{\bf E,S0} & {1} & $T\leq 0.5$ & 21.0\\ 
{\bf Sa} & {2} & $0.5<T\leq 2.5$ & 16.9\\ 
{\bf Sb} & {3} & $2.5<T\leq 4.5$ & 20.9\\
{\bf Scd} & {4} & $4.5<T\leq 8.5$ & 30.1\\ 
{\bf Irr} & {5} & $8.5<T$ & 11.1\\ 
\end{tabular}
\end{table}
The Normal26 spectra have been split into five broad groups, based on
their visual morphology. These groups can be seen in Table 1 which
also gives the percentage of galaxies falling into each group for the
ESO catalogue (Lauberts \& Valentijn 1989).  We decided to bin the
data in this way so that there are a number of spectra in each group
and the ANN is not over-trained to recognize a specific spectrum for a
particular class. With the small data set that we have, this is still
a risk, but the combination of this binning and using many noisy
deviates of each spectrum helps to alleviate the problem.
Unfortunately the `Irr' group is not well represented in the Normal26
sample, since all but two of these galaxies fall into the Unusual29
set. 
\begin{figure}
\label{fig1}
\psfig{figure=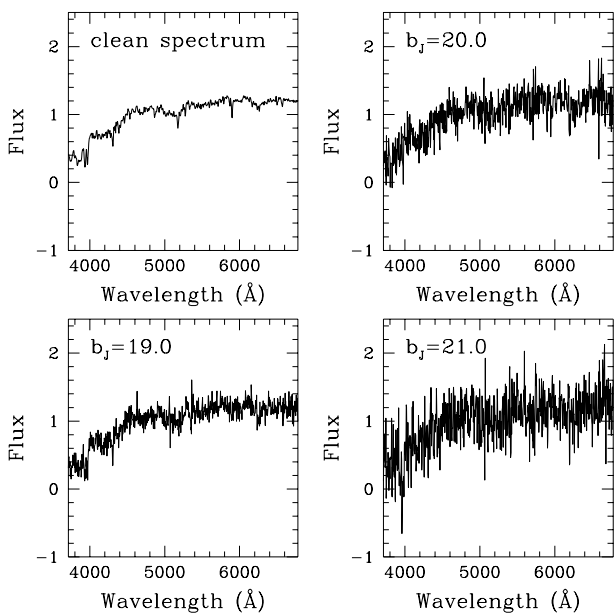,width=3.4in,height=3.4in}
\caption{Simulated spectra based on NGC 3379 (E0).}
\end{figure}
The Unusual29 spectra have also been tentatively placed in these
5 groups, based on their visual morphology where it is
defined. Spectra without a defined morphology, or purely labeled as
peculiar, have been placed in the final bin.  Table 2 summarizes the
data set. A notes section is also given for the Unusual29 spectra,
which specifies why they have been categorized as unusual.  Two of the
spectra were found to adversely bias the PCA, and so have been removed
from the normal set of spectra. In the case of NGC 1569 this is due to
serious galactic reddening being evident in the continuum. In the case
of MK 487 the reason is less obvious, but visual inspection of the
spectrum (Kennicutt 1992) indicates an erratic continuum.

\begin{table*}
\caption{The selection of galaxy spectra from Kennicutt (1992).}
\begin{tabular}{lllp{1cm}llll}
\hline
\multicolumn{3}{c}{\bf Normal26} && \multicolumn{4}{c}{\bf Unusual29} \\ \hline
\hline
{\em Galaxy} & {\em Morphology} & {\em Group} && {\em Galaxy} & {\em Morphology} & {\em Group} & {\em Notes}\\ \hline
\hline

NGC3379   &  E0    &  E,S0   &&   MK487     &  Im    &  Irr  &  Odd (see text) \\                  
NGC4472   &  E1/S0 &  E,S0   &&   NGC1569   &  Sm/Im &  Irr  &  Reddened \\                    
NGC4648   &  E3    &  E,S0   &&   NGC4670 &   SB pec  & Irr & Peculiar \\              
NGC4889   &  E4   &   E,S0  &&    NGC3034 &   I0      & Irr & Peculiar\\               
NGC3245   &  S0    &   E,S0  &&   NGC3077 &   I0      & Irr & Peculiar\\               
NGC3941   &  SB0/a &   E,S0  &&   NGC5195 &   I0 pec  & Irr & Peculiar\\              
NGC4262   &  SB0   &    E,S0 &&   NGC6240 &   I0 pec  & Irr & Peculiar\\               
NGC5866   &  S0    &   E,S0  &&   NGC3310 &   Sbc pec & Sb & Global Starburst\\        
NGC1357   &  Sa    &   Sa &&      NGC3690 &   Sc pec  & Scd & Global Starburst\\       
NGC2775   &  Sa    &  Sa  &&      NGC6052 &   Sm pec  & Irr & Global Starburst\\      
NGC3368   &  Sab   &  Sa  &&      UGC6697 &   S pec   & Irr & Global Starburst\\      
NGC3623   &  Sa    &  Sa  &&      NGC2798 &   Sa pec  & Sa & Starburst Nucleus\\        
NGC1832   &  SBb   &  Sb  &&      NGC3471 &   Sa      & Sa & Starburst Nucleus\\        
NGC3147   &  Sb    &  Sb  &&      NGC5996 &   SBd     & Irr & Starburst Nucleus\\       
NGC3627   &  Sb    &  Sb  &&      NGC7714 &   S pec   & Irr & Starburst Nucleus\\       
NGC4750   &  Sbpec &  Sb  &&      MK35    &   pec     & Irr & Peculiar\\               
NGC2276   &  Sc    &  Scd  &&     MK59    &   SBm/Im  & Irr & HII Region\\              
NGC4775   &  Sc    &  Scd  &&     MK71    &   SBm     & Irr & HII Region\\            
NGC5248   &  Sbc   &  Sb  &&      NGC3516 &   S0      & E,S0 & Seyfert I\\            
NGC6217   &  SBbc  &  Sb  &&      NGC5548 &   Sa      & Sa & Seyfert I\\              
NGC2903   &  Sc    &  Scd  &&     NGC7469 &   Sa      & Sa & Seyfert I\\              
NGC4631   &  Sc    &  Scd  &&     NGC3227 &   Sb      & Sb & Seyfert II\\               
NGC6181   &  Sc    &  Scd  &&     NGC6764 &   SBb     & Sb & Seyfert II\\               
NGC6643   &  Sc    &  Scd  &&     MK3     &   S0      & E,S0 & Seyfert II\\             
NGC4449   &  Sm/Im &  Irr  &&     MK270   &   S0      & E,S0 & Seyfert II\\             
NGC4485   &  Sm/Im &  Irr  &&     NGC1275 &   E pec   & E,S0 & Peculiar\\              
	  &        &	   &&     NGC3303 &   pec     & Irr & Peculiar\\          
	  &        &       && 	  NGC3921 &   S0 pec  & E,S0 & Peculiar\\         
	  &        &       &&  	  NGC4194 &   Sm pec  & Irr & Peculiar\\

\end{tabular}
\end{table*}
\begin{figure}
\label{fig2}
\psfig{figure=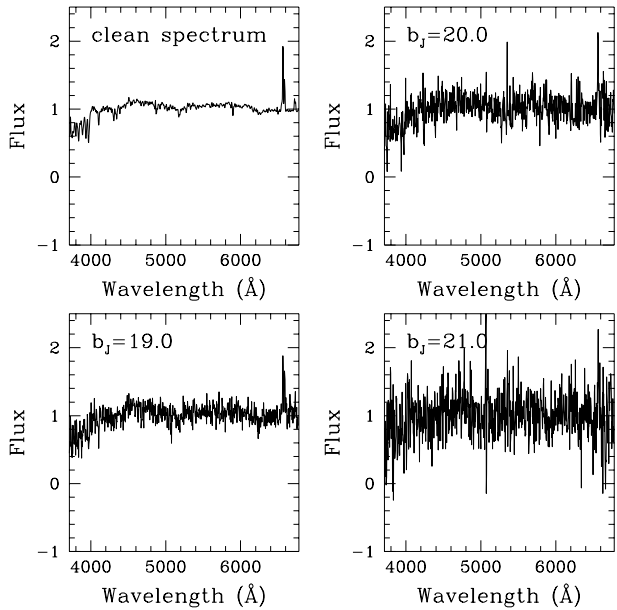,width=3.4in,height=3.4in}
\caption{Simulated spectra based on NGC 3627 (Sb).}
\end{figure}
In order to accurately test the methods for spectral reconstruction
and classification it is first necessary to produce a set of galaxy
spectra which resembles the spectra received from large redshift
surveys. Details for the 2dF system (Taylor 1994) on the
Anglo-Australian Telescope are used to degrade the Normal26 spectra,
to simulate spectra from objects with a range of $b_{\rm J}$
magnitudes. Appendix A gives details about the spectral simulation
procedure, detailing how the system response function, sky spectrum,
fibre size and galaxy magnitude are incorporated into the
simulations. It should be noted that it is difficult to predict the
exact performance of the 2dF system and observations will obviously
differ due to conditions, hence the simulations remain approximate,
but demonstrate the level and variation of noise across the
spectrum. We do not deal with the effect of aperture bias (the fact
that a fibre can only sample a small area of a bright galaxy) in this
paper, but acknowledge that it may cause discrepancies between the
morphological and spectral type determined for a galaxy. Zaritsky et
al. (1995) find that in general aperture bias would not constitute a
large effect for the majority of galaxies, but they stress that it may
still pose a problem in some cases. Figures 1-3 show examples of the
simulated spectra for an elliptical galaxy, a spiral galaxy and an
irregular emission line galaxy. In each case the original spectrum and
simulations at a $b_{\rm J}$ magnitude of 19, 20 and 21 are shown.  We
refer to the original spectra (from Kennicutt 1992) in the following
sections as the `clean' spectra, and we consider them not to contain
noise. This is a reasonable assumption, since figures 1-3 show that
the low S/N in the noisy simulations makes any noise in the original
spectra negligible. Figures 1-3 indicate the importance of the
emission lines for spectral classification at low S/N, and also reveal
how the additional noise due to sky lines can produce false features
such as those seen in figure 2, at $b_{\rm J}=20$ and $b_{\rm J}=21$.

For this method, the spectra must be compared in their rest frame, so
we have used the redshifts (from Kennicutt 1992) for the galaxies to
de-redshift the spectra (see Appendix A for the full procedure). The
accuracy of the redshifts and resolution of the spectra determine the
number of wavelength points which are used for the PCA.

\begin{figure}
\label{fig3}
\psfig{figure=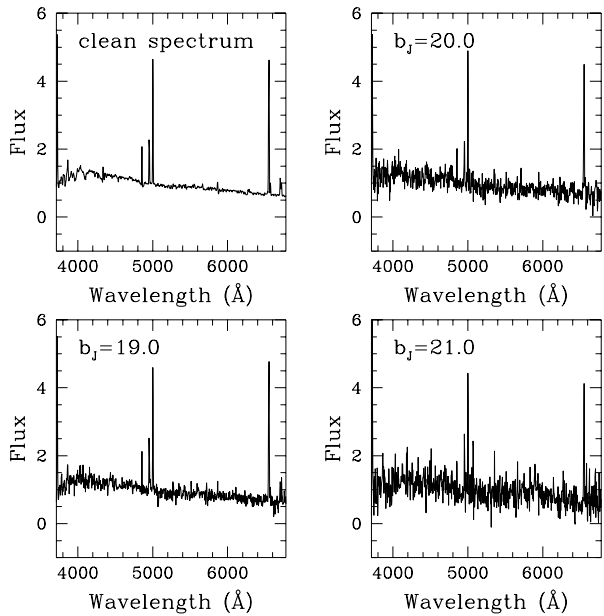,width=3.4in,height=3.4in}
\caption{Simulated spectra based on NGC 4485 (Im).}
\end{figure}
\begin{figure}
\label{fig4}
\psfig{figure=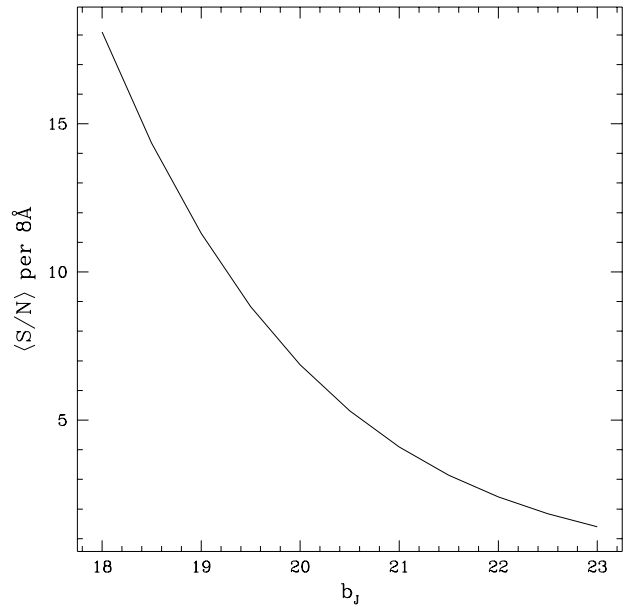,width=3.4in,height=3.4in}
\caption{The variation in signal to noise per 8\AA \ resolution
element measured for simulated spectra as a function of $b_{\rm J}$.}
\end{figure}

A set of 900 spectra are produced in this way for each $b_{\rm J}$
magnitude. These 900 are based upon the Normal26 spectra, but each
spectrum is simulated $N_{\rm s}$ times, where $N_{\rm s}$ is selected
so that the final 900 spectra have the same morphological distribution
as the ESO catalogue, as given in Table 1.

Since our initial data set is limited, this set of 900 spectra does
not contain all the variation in a true observed set of spectra, such
that we acknowledge that this analysis may lead to an optimistic rate
of classification, but it demonstrates the methods we wish to use on
more extensive sets of observed spectra.

The S/N per 8\AA \ resolution element averaged over all wavelengths
for all 900 spectra is calculated when a set of spectra is produced
for a given $b_{\rm J}$.  Figure 4 shows a plot of this $\langle S/N
\rangle$ against $b_{\rm J}$ computed in this way which can be used to
associate the $b_{\rm J}$ magnitudes used in this paper with a general
S/N spectrum from any source.

\section[]{Principal Component Analysis}
Principal Component Analysis (hereafter PCA) is a technique for both
data compression and analysis (Murtagh \& Heck 1987) which can be
employed to assess variations in galaxy spectra. By identifying the
linear combination of input parameters with maximum variance, a set of
new axes (principal components) is derived. A mathematical description
is given in Appendix B. Computationally, we use the technique of
Singular Value Decomposition to find the eigenvectors (or principal
components) of the covariance matrix.

The question which arises is how to create the ideal set of principal
components (or PCs) for galaxy spectra. The Normal26 spectra provide a
useful data set, but they are not a representative sample of galaxy
spectra in general.  Ideally it would be best to define the PCs from
the observed galaxy spectra of a large survey, but this data would be
noisy and it is not clear at first how the noise would affect the
PCA. Possibly filtering the spectra to reduce the noise level could
improve the analysis, but this will lead to a loss of information.

As explained in section 2, a set of 900 clean spectra are created,
based upon the Normal26 galaxies , which are sampled in 4\AA \ bins
resulting in 768 wavelength bins for each spectrum. PCA is now
conducted on the (768x768) covariance matrix using the techniques
outlined in Appendix B. It should be noted that the speed of the PCA
algorithm is dependent only on the number of wavelength bins used and
not on the number of spectra (we chose 900 purely to produce a large
morphologically weighted data set with many random variations).

\subsection[]{Variance Scaling and normalization}
In the analysis which follows, the spectra are all normalized to have
the same total flux over the wavelength range considered, then the
mean spectrum is subtracted from each of the input spectra. No further
scaling is used.

We did examine the possibility of scaling each input flux to
unit-variance across the sample of spectra. This method is sometimes
recommended for PCA analyses since it places each input on an equal
footing. This would be advantageous when considering object attributes
which are fundamentally different, for example if we were basing a
classification system on galaxy colour, image ellipticity and OII
equivalent width. For spectra, the problem is slightly different since
all the inputs are fluxes for different wavelengths, hence the
relative strengths of the inputs is important and should be retained
in the analysis. Francis et al. (1992) investigate this problem for
PCA on QSO spectra and choose not to scale by the variance.

Scaling by the variance means that the PCs for galaxy spectra at
different noise levels are radically different, since at high S/N the
PCA is sensitive to well correlated small features, but at low S/N
these features are lost. Having chosen not to use this scaling we find
the PCs only vary slightly with noise level, hence the PCs are
intrinsic to the galaxy spectra themselves and not to particular
observing conditions. In this case the PCs relate to both the
magnitude and correlation of the features being chiefly concerned with
regions of the spectra where the signal is strongest, such that they
are not swamped by noise in the continuum. We also did the analysis
with variance scaling, but found it gave a less reliable final
classification, so we do not discuss this further.

Other possibilities are a normalization of each spectrum such that the
integrated flux is the same, or alternatively such that the sums of
the squares of the fluxes across the spectrum is unity (unit scalar
product). This second case has certain mathematical benefits since it
means that each spectrum can be represented by a unit vector in the
parameter space. Connolly et al. (1995) consider these possibilities,
but they find their results are not greatly affected by the choice of
normalization. Hence for simplicity, we opt purely for a normalization
to equal integrated flux.

The other operation we perform is to subtract the mean spectrum from
each of the spectra in the set. This centres the points in the PC
space about the origin, and makes the PCs easier to interpret.

\subsection[]{The meaning of the principal components}
\begin{figure}
\label{fig5}
\psfig{figure=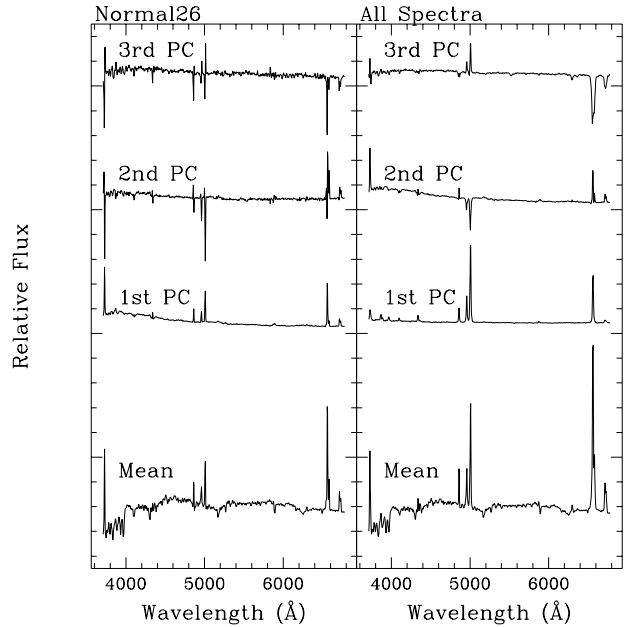,width=3.4in,height=3.4in}
\caption{Principal components for the Normal26 spectra and for the
entire set, without additional noise.}
\end{figure}
Figure 5 shows the mean and the first 3 PCs for the data set based
upon the Normal26 spectra and Figure 6 shows an enlargement of the
first PC indicating the important features. These are computed without
noise being added to the spectra, but when noise is added, the PC axes
change very little. We quantify this affect in more detail in section
4. We find that the first PC accounts for 87\% of the variance in the
set of 900 clean spectra based on the Normal26 spectra. When the set
is simulated at $b_{\rm J}=22$ $(S/N\sim2.5)$ the plot of the first PC
is qualitatively the same, but only accounts for 11\% of the variance,
since the noise is producing large amounts of uncorrelated
variance. Let us consider the meaning of the correlations which have
been found in the Normal26 spectra. It can be seen that the first PC
represents the correlation in the strength of the emission lines with
the young stellar component.  It shows that the OII (3727), OIII (4959
and 5007), $H\alpha$ (6563) and $H\beta$ (4861) lines are all linked
with a blue continuum, demonstrating the effect of the ionizing
photons from young stars exciting the interstellar medium resulting in
strong emission lines. The second PC allows for a range of ionization
levels of the galaxies, since the oxygen and hydrogen lines are
anti-correlated. The third PC indicates numerous other correlations
between absorption and emission features. A parallel study involving
PCA with the Kennicutt sample of galaxies (Sodr\'e \& Cuevas 1996)
finds similar principal components.

\begin{figure}
\label{fig6}
\psfig{figure=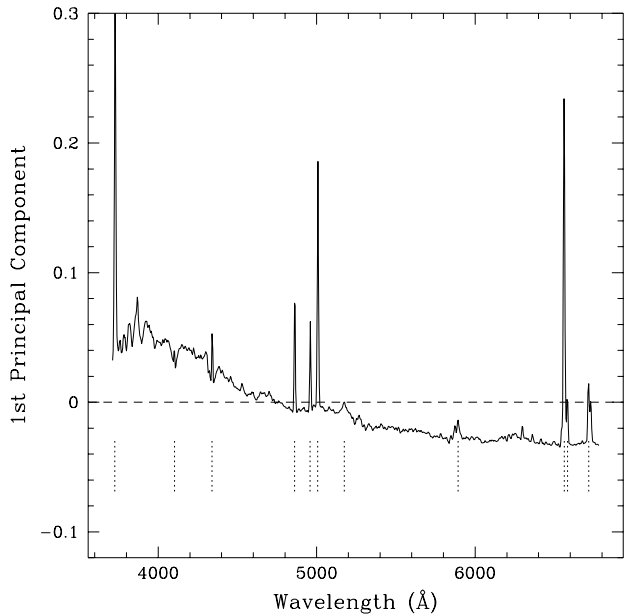,width=3.4in,height=3.4in}
\caption{The first principal component for the Normal26 spectra,
indicating some of the important features.}
\end{figure}
\begin{figure}
\label{fig7}
\psfig{figure=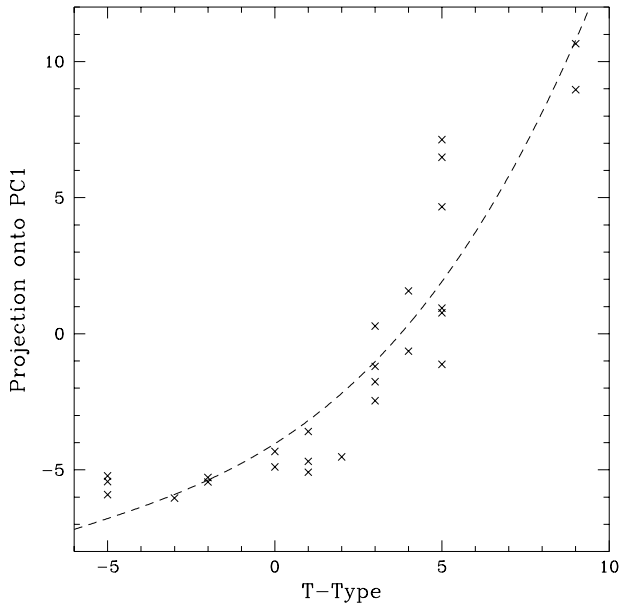,width=3.4in,height=3.4in}
\caption{The projection onto the first principal component plotted against 
T-Type for the Normal26 spectra.}
\end{figure}
\begin{figure}
\label{fig8}
\psfig{figure=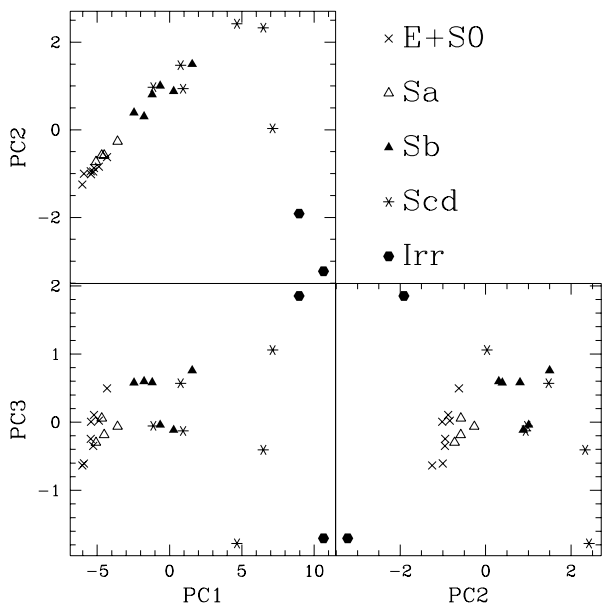,width=3.4in,height=3.4in}
\caption{Projections onto the first 3PCs plotted against one another
for the Normal26 spectra.}
\end{figure}

It is hoped that the amplitude of a small number of these eigenspectra
in any particular spectrum will be sufficient to spectrally classify
the galaxies. Since the projections onto the eigenspectra use
information from the entire spectrum this provides a much less noisy
measure than comparing the strength of particular features. A
simple indication that this may be true can be seen in Figure 7, which
indicates a correlation between the projection (see Appendix B) of
each of the Normal26 spectra onto the first PC and the morphological
type of the galaxy on the T-Type system (de Vaucouleurs 1959; de
Vaucouleurs 1963). A simple fit to this relation (as shown by the
dotted line) allows the projection onto the first PC to be used to
classify the galaxies into a specific T-Type. This has been used later
to provide a comparison to the classifications using the ANN. Further
evidence of an underlying spectral sequence can be seen in Figure 8
where the projections onto the first three PCs are plotted against
each other, with the symbols representing different morphological
types. Segregation in this plot indicates that the PCs are capable of
differentiating between galaxy morphologies and that the Hubble
sequence is clearly evident in the Normal26 spectra and not just their
visual appearance on the sky. The Hubble sequence appears as a
combination of PC1 and PC2, indicating that although one spectral
parameter is sufficient to explain the sequence up to Sc galaxies, a
second parameter (allowing for variation in ionization) becomes very
important for late type and irregular galaxies, where strong star
formation leads to high ionization levels. In their study, Sodr\'e
\& Cuevas (1996) found that stellar synthesis models (Bruzual \&
Charlot 1995) with different ages and star formation rates form a
similar sequence when projected onto PCs derived from the Kennicutt
(1992) sample of galaxies.

Figure 5 also shows the mean and first three PCs for the entire data
set, including the Normal26 and the Unusual29 spectra. These are
derived by the same method as described above. The morphological mix
from Table 1 is again used, but note that this does not necessarily
represent a true spectral mix. This means that the galaxies with
unusual features are over represented, but this allows their affect on
the PCA to be seen. In this case, the first PC is entirely due to the
emission line strength, since the PCA is dominated by the emission
line objects. The young stellar continuum is evident in the second PC
along with anti-correlations between the major emission lines. In this
respect the first PC from the Normal26 spectra is evident as a
combination of the first and second PCs of the entire data set.  The
third PC is now quite different, and displays the broad hydrogen lines
characteristic of the galaxies with Seyfert nuclei.

\section[]{Reconstructions from noisy spectra}
We now proceed to investigate the effect of noise on the PCA
technique. It would be useful to perform PCA on a large set of
observed spectra, such as the 2dF spectra, since this set would
contain all the possible variation in galaxy spectra and is
representative of the local universe. However these observed spectra
would be noisy and this may affect the location of the
principal components. Using the spectral simulations we can assess the
nature of this effect, by measuring the ability of the PCs to
reconstruct the original spectrum. Appendix B explains how a
reconstruction of the original data can be found using only a small
number of PCs. Taking a set of noisy simulated spectra we can
reconstruct them and define the total Residue R for the set as
$$
R={1 \over NM}\sum_{ij}(S^{r}_{ij}-S_{ij})^2 ,       \eqno (1)
$$
where the sum is over all $N$ spectra and $M$ wavelengths, $S^{r}_{ij}$ is
the flux of the ith reconstructed spectrum at wavelength j, and
$S_{ij}$ is the flux of the original clean ith spectrum at wavelength
j. This reconstruction technique suggests the useful ability of PCA to
disregard the noise, which is assumed to be uncorrelated. The major
correlations in the signal are selected by the PCA, and the noise only
interferes with the later PCs, such that a reconstruction using the
most significant PCs can eliminate much of this uncorrelated noise. To
demonstrate this effect, figure 9 shows a plot of R against the number
of PCs used in the reconstruction. Seven different cases are
considered and these are listed in Table 3. In each case two sets of
spectra are used. One set is used to define the principal components
and is labeled `Spectra$_{\rm PCA}$'. The other set is reconstructed
to find the Residue R and is labeled `Spectra$_{\rm REC}$'. This
reflects the possibility of defining the principal components prior to
the observations using a smaller set of high quality spectra. However
this would probably entail using a limited data set, so Table 3 also
contains a case where only half of the Normal26 spectra have been used
to define the PCs. Each set contains 900 spectra mimicking the
morphological mix given in Table 1.  There is one other possibility
considered in Table 3, that of filtering the spectra prior to the
reconstruction and we have used the technique of Wiener filtering to
demonstrate this effect (see section 4.2).

\subsection[]{Reconstruction results}
Referring to Figure 9, we can see that the smallest R value is
obtained using the noise free spectra for the Spectra$_{\rm PCA}$ and
the Spectra$_{\rm REC}$ sets (line a). This is to be expected and
represents the ideal condition, but one which is not available for a
real observation since the underlying signal is not known. A set of
clean spectra, such as the spectra we are using, could be used to form
the PCs for use with observed noisy spectra, as lines b and e
demonstrate. However the line d indicates that when only half of the
spectra are used in the PCA the result is considerably worse, which
suggests that a small set of spectra should not be considered
representative of a larger ensemble. Therefore it would be better to
derive the PCs from the noisy spectra themselves. This condition is
shown by lines c and g for two different noise levels. It can be seen
that for $b_{\rm J}$=19 (line c) about 8 PCs still contain meaningful
information, but the later PCs are merely reconstructing the noise (as
indicated by a rise in R). This means that the optimal reconstruction
is found by limiting the PCs used to the point at which R is found to
be a minimum and this then represents the entire meaningful
information that can be extracted from the spectrum. Line g indicates
that for very noisy data only the first PC is still meaningful.

\begin{table}
\caption{The seven PCA combinations used for analysis of reconstruction errors, with the notation for Figure 9.}
\begin{tabular}{llc}
\hline
Spectra$_{\rm PCA}$ & Spectra$_{\rm REC}$ & Notation in Figure 9\\ 
\hline
\hline
Clean & Clean  & a\\ 
Clean & $b_{\rm J}=19$ & b\\
$b_{\rm J}=19$ & $b_{\rm J}=19$  & c\\
Clean, half data set & Clean & d\\ 
Clean & $b_{\rm J}=22$ & e\\ 
$b_{\rm J}=22$ & $b_{\rm J}=22$ filtered  & f\\
$b_{\rm J}=22$ & $b_{\rm J}=22$  & g\\ 

\end{tabular}
\end{table}
\begin{figure}
\label{fig9}
\psfig{figure=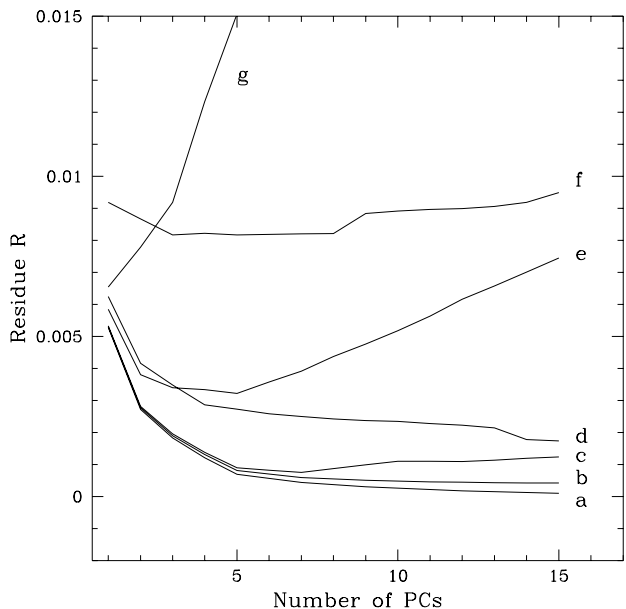,width=3.4in,height=3.4in}
\caption{Reconstruction errors for different sets of spectra. See
Table 3 for explanation of line labels.}
\end{figure}
\begin{figure}
\label{fig10}
\psfig{figure=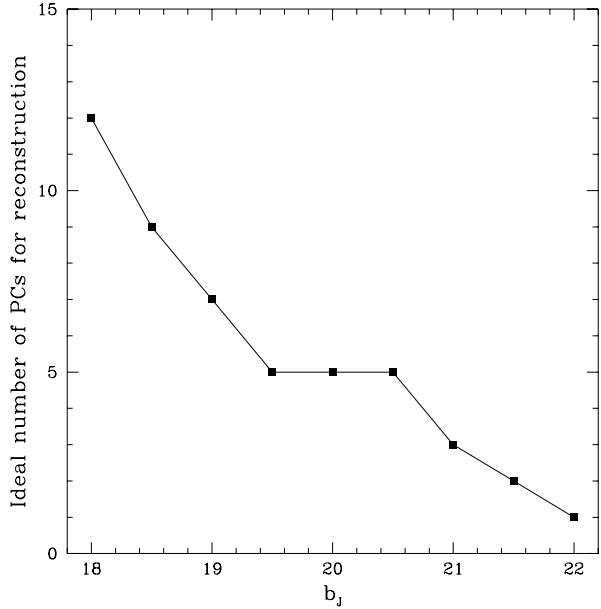,width=3.4in,height=3.4in}
\caption{The optimal number of PCs for spectral reconstruction as a
function of $b_{\rm J}$.}
\end{figure}
So given a set of noisy spectra, it is reasonable to perform PCA on
the spectra themselves, but to acknowledge the fact that only a
certain number of the PCs are useful, with this number depending on
the S/N of the spectra. If a set of high S/N spectra are available,
using the PCs from these may extract more of the information (see
lines e and g), but in this case the assumption that the noisy spectra
can be well described by the PCs from the clean spectra must be made,
and (as line d shows) this is not always true. Figure 10 indicates how
the optimal number of PCs (i.e. the number which gives a minimum in R)
varies with $b_{\rm J}$ when Spectra$_{\rm PCA}$ and Spectra$_{\rm
REC}$ are both simulated at $b_{\rm J}$. The exact normalization of
this graph depends on the specific data set being considered,
including factors such as the number of spectra, and the wavelength
sampling.  For a very large set of data from a big redshift survey it
may be expected that more PCs would be significant.

\subsection[]{Wiener Filtering}
An alternative method for extracting the meaningful information from
the spectra is Wiener filtering in Fourier space (see Press et
al. 1992 for a full description). This involves a smooth truncation of
modes in a data independent basis, as opposed to the PCA which
involves a sharp truncation in a basis which is adapted to the
data. In Fourier space, let S(k) be the true spectrum of a galaxy,
then the observed spectrum O(k) is given by

$$O(k)=S(k)+N(k) , \eqno (2) $$
where $N(k)$ is the Fourier transform of $n(\lambda)$ (the noise
at each wavelength). Let us define a linear filter in Fourier space, W(k) by

$$S_{\rm r}(k)=O(k)W(k) , \eqno (3) $$
where $S_{r}(k)$ is the best reconstruction of S(k). By a least squares
minimization with respect to $W(k)$ we find

$$W(k)=\frac{|S(k)|^2}{|S(k)|^2+|N(k)|^2} , \eqno (4) $$ 
where we have ignored terms involving $S(k)N(k)$ since the noise and
signal are considered to be uncorrelated. From equation (4), we define
$W(k)$ as the Wiener filter.

For this method we must first assume an underlying signal $S(k)$ in
the spectrum. To do this we have formed 5 templates, one for each of
the groups given in Table 1 by taking the mean of the Normal26 spectra
in that class. We then take each noisy spectrum and compare it to the
5 templates and look for the best template in the least squares
sense. This template is then used as the prior for the Wiener
filtering. In addition this template matching method is a simple
method of classification where a galaxy spectrum can be allocated to
the group whose template it matches best. We use this later as a
comparison to the classifications from the ANN.

Wiener filtering can be seen as an alternative technique to produce a
reconstruction from a noisy spectrum and it is interesting to compare
the PCA reconstructions and the Wiener reconstructions. This
comparison can be seen in Figure 11 for different levels of noise. It
can be seen that the Wiener filtering reduces the noise, but also
smoothes the signal, such that at low S/N the features are lost and
only the rough continuum shape remains. In comparison the PCA
reconstructions retain much more of the information in the spectrum,
producing a reasonable reconstruction of the spectrum to a $b_{\rm J}$
of 22. At low S/N the noise causes some spurious effects, but many of
the distinguishing spectral features, in this case $H\alpha$ and the
4000\AA \ break are retained. In order to make this reconstruction,
the noisy spectrum is assumed to be characteristic of the set of
spectra used to produce the principal components. This means that the
reconstruction conforms to the correlations laid by the PCA. For some
data sets it is possible that PCA would not provide a good description
of the data and Wiener filtering in Fourier space would be the
prefered method. Ideally a very large set of spectra is required for
the PCA, such that the complete range of spectral possibilities is
encompassed, and we hope to apply the techniques given here to such
data sets in the future.
\begin{figure*}
\label{fig11}
\psfig{figure=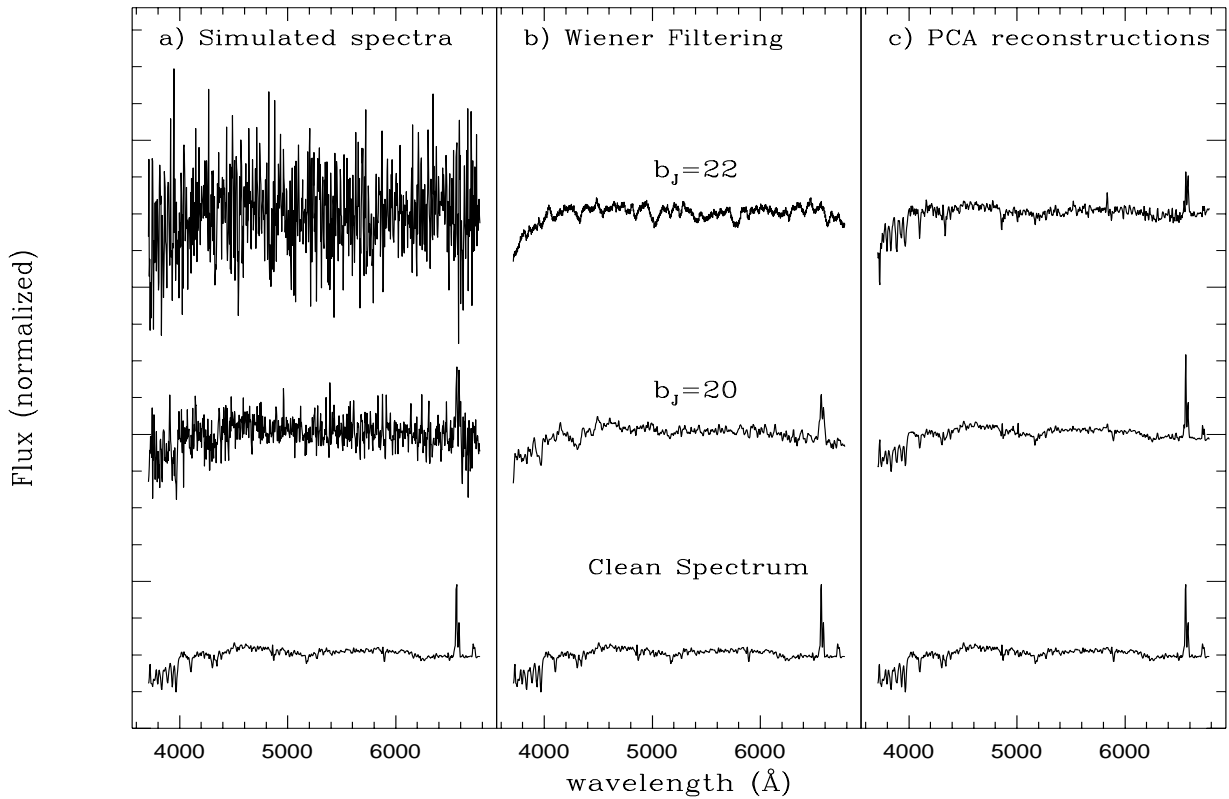,width=7.0in,height=4.5in,angle=270}
\caption{Comparison of spectral reconstructions for NGC3627 (Sb). a)
The simulations for $b_{\rm J}$=20 and $b_{\rm J}$=22. b) Wiener
filtering of the noisy spectra using a group template (see text). c)
PCA reconstructions of the noisy spectra based upon 8PCs derived from
the Normal26 spectra.}
\end{figure*}
Referring back to figure 9, line f is the result of first Wiener
filtering the spectra before projecting onto the PCs. This removes
much of the noise so that the line does not rise so rapidly, but the
action of the filtering also removes much of the meaningful signal in
the spectrum so that the reconstruction is never as good as the single
PC reconstruction based on the noisy data (line g).
\subsection[]{Combining Wiener filtering and PCA} 
As an aside, we can see in the previous section that the PCA, which
takes into account correlations between the features, produces a far
superior reconstruction than the Wiener filter used in Fourier space,
but we have also noted that PCA reconstructions of noisy data should
be restricted to only the first few PCs, since the noise interferes
with the later PCs. The PCA works better than the Fourier
representation because the PCA axes are chosen specifically to
represent the data, whereas the Fourier axes are a generalized
orthogonal set and not specific to the data being considered. The
Wiener filter is used to produce a smooth cutoff of the Fourier modes
so that the noisy modes are reduced in weight. Such a procedure could
also be used with the PCA where, instead of truncating after a
determined number of PCs, a filter is used which merely reduces the
weight of the later PCs. In this paper, we actually need a direct
truncation of the PCs, since we want to minimize the number of inputs
to the ANN (hence reducing the number of free parameters of the
network), but for a general spectral reconstruction the filtered PCA
is a promising idea.

\section[]{Spectral classification with an artificial neural network}

Figures 7 and 8 show that the visual morphology and the spectrum of a
galaxy are related and that this relation is embodied in the
projection onto the PCs.  This suggests that a useful method for the
classification of galaxy spectra is to associate each spectrum with a
morphological type.  This would allow the morphology of galaxies to be
examined at far greater distances and with less subjectivity than
conventional examination of galaxy images. We have trained an ANN to
assign morphological classifications to galaxies, based on their
spectra as represented by the projections onto a small number of PCs.

For each spectrum, the ANN produces an output classification which is
a non linear function of the inputs. The form of the function is
parameterized by a set of weights which are adjusted so that the
output matches the known classifications for a training set.  To be
precise, the effect of training the ANN is to perform a minimization
across the ANN weights vector ({\bf w}) given a set of inputs ${\bf
x}_i$ for the ith galaxy (e.g. the spectrum as represented by the PCs)
and known outputs $T_{\rm i}$ (e.g. the morphological group). This is
done by minimizing the cost function
$$
E = {1 \over 2} \sum_i [T_i - F({\bf w}, {\bf x}_i) ] ^2, \eqno (5)
$$
where the non-linear function $F({\bf w}, {\bf x})$ represents the
network and the summation is over the training set of spectra.
\begin{figure}
\label{fig12}
\psfig{figure=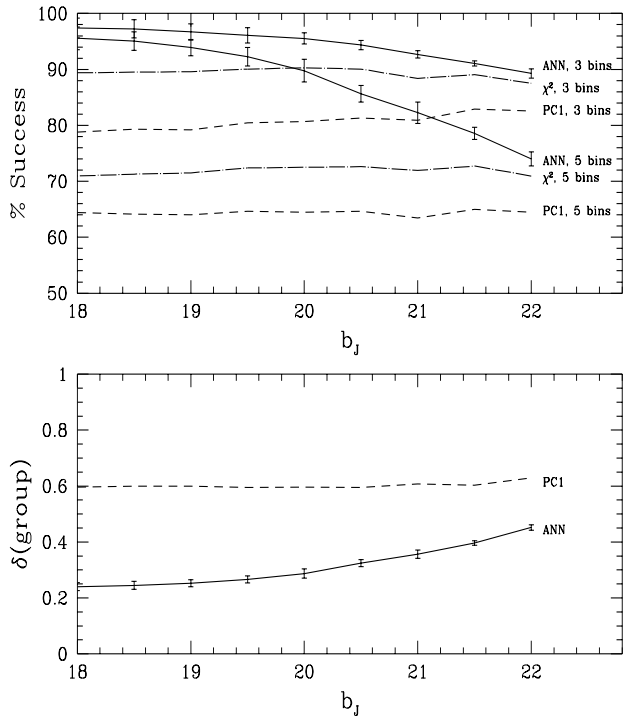,width=3.4in,height=4.0in}
\caption{The percentage of ANN classifications which agree with the
known morphological types for 5 and 3 classes, based upon the Normal26
spectra. Also shown is the success of $\chi^{2}$ template matching and
classification based solely on PC1. The lower graph indicates the
variation in $\delta(group)$ against $b_{\rm J}$ for classifications from
the ANN, and PC1 alone.}
\end{figure}

Once the weights are set, the training is complete and the ANN can be
used to classify the complete galaxy sample. A full description of the
ANN as a tool for data analysis is given in Appendix C (for further
detail see Lahav et al. 1996). We used a quasi-newton ANN code with
the network architecture designed to allow the projections onto the
first 8 PCs (derived from a set of spectra simulated at $b_{\rm
J}=19.0$) to be used as input to the net and a single output being the
morphological group. Between the input and output layers we chose a
single hidden layer with 5 nodes, which provides a level of
non-linearity in the classification. We experimented with
different numbers of nodes and hidden layers and decided upon the
8:5:1 setup as the simplest architecture which gave consistently
successful results. Simpler architectures were not reliable and more
complex nets failed to improve the results. The output from the
ANN could then be scaled and binned to give the five classes as
defined in Table 1. The training process involved weight decay, which
acts as a regularisation during the training, preventing erratic
variations in the weights. The quasi-newton minimization and the use
of weight decay are discussed in more detail in Appendix C.

We chose to use 8 PCs based upon the results in section 4, which
indicate this to be a reasonable compression of the data for the S/N
levels we considered. We produced a set of 900 simulated spectra at
each of 9 values of $b_{\rm J}$ between 18 and 22, resulting in a total
set of 8100 spectra. One third of this set was then repeatedly
submitted to the ANN as a training set until the error between the ANN
classification and the known morphological types of the galaxies began
to converge. The `trained' net was then used to classify the
complete set of 8100 spectra onto a continuous scale defined by the
group G number as given in Table 1. In this way the scaled output from
the ANN was a single number in the range 0.5 to 5.5 and an output
group was found by allocating the galaxy to the nearest group
bin. These classifications could be compared with the known types of
the galaxies to give a level of success at each magnitude. The ANN is
trained, and spectra classified, ten times using this method to give a
mean and standard deviation for the percentage of galaxies allocated
to the correct group, and these results can be seen in Figure 12. In
addition, if only a 3 group classification is required, the Sa, Sb and
Scd groups can be combined to give one large group of spirals. The
success rate for this 3 group binning is also shown in figure 12.

A further measure of success is the $\delta(group)$ statistic given by

$$\delta(group)=\sqrt{{1 \over N}{\sum_{i}(G_{\rm i}-A_{\rm i})^2}} , \eqno
(6) $$ where the sum is over the N spectra (in this case a set of
900), $G_{\rm i}$ is the actual group to which the galaxy belongs, as
defined in table 1 and $A_{\rm i}$ is the neural net classification on
a continuous scale from 0.5 to 5.5.

These results are encouraging, showing that the morphological
variation in the Normal26 galaxies is well represented in their
spectra and that the PCA/ANN technique is capable of extracting this
information even with very noisy spectra.

Figure 12 indicates several other lines for comparison. Two of the
lines refer to the classification success when a classification based
solely upon PC1 is used. A simple relation between T-Type and PC1 is
assumed (as indicated by the line on Figure 7) and the results shown
using 5 and 3 groups. $\delta(group)$ was also calculated for this
method so that it can be compared with the ANN. The classification
based solely on PC1 is found not to be ideal, although it is stable to
high noise levels (since PC1 is the most meaningful correlation in the
data and should not be greatly affected by noise). It is clear that
the ANN is capable of better classifications using more of the
principal components than the single PC result. The later components
are affected by noise to a greater extent, so the ANN classification
does fall as the noise level increases.

The other two lines on Figure 12 refer to a classification based
on $\chi^2$ template matching, from the procedure in section 4.2. This
indicates a reasonable level of success, but is unable to capitalize
on the extra information at high S/N which allows the ANN to refine
the classifications. Since the template matching gives a discrete
group output, $\delta(group)$ is not calculated for this method.

\section[]{Agreement of morphological and spectral types}
We now have an ANN which has been trained to relate the spectral type
and morphological type of normal galaxies, using the projections onto
the first eight PCs (derived from the Normal26 morphologically
weighted sample). We use this to classify the Normal26 and the
Unusual29 spectra without noise, to gain an indication of the
agreement between spectral and morphological type. The results can be
seen in figure 13. The Normal26 spectra form a reasonable sequence,
with a degree of scatter in each group. This scatter is related to the
$\delta(group)$ statistic plotted in Figure 12 (but note that
$\delta(group)$ is summed over a complete morphological sample of 900
spectra at a particular noise level). Some overlap between the groups
can be seen, verifying the fact that we are dealing with a sequence in
galaxy type and not discrete classes. The agreement between
morphological and spectral type indicated in Figure 13 substantiates
the conclusions of the PCA analysis (Figures 7 and 8) which suggested
strong links between spectra and morphology. It is reassuring to see
that the traditional Hubble classification system is telling us about
stellar and gas content in addition to the morphology of the
galaxy.

As expected, the Unusual29 spectra do not conform to this
morphology-spectrum relation. In general, the unusual spectra which
have been morphological classified into groups 1 to 4, produce a
higher spectral class from the ANN. This is due to the presence of
starbursts, Seyfert nuclei and emission features which increase the
`activity' in these spectra, above that of a normal galaxy for that
class. The unusual spectra in morphological group 5 contain galaxies
with T-Types greater than 8.5, which include irregular and peculiar
types along with extreme emission galaxies. The irregular emission
line objects are classified correctly as being extreme in spectral
class (group 5), but the peculiar galaxies reveal a range of spectral
features, such that they are classified into a variety of spectral
classes. We can investigate some particular cases which have been
highlighted in Figure 13, and look at the spectra and comments given
in Kennicutt (1992). The morphologically peculiar galaxy NGC 3077 has
been spectrally classified by the ANN as a late Sb galaxy (ANN output
group 3.15) which broadly agrees with the comments given in Kennicutt
(1992) that this spectrum is similar to that of a normal Sc. In
contrast, the spectrum of NGC 5195 has been spectrally classified as
an elliptical by the ANN (ANN output group 1.25), whereas Kennicutt
(1992) comments that it resembles an old stellar population with weak
emission lines, or an `E+A' galaxy. From a sample of this size, we
cannot say that the ANN is telling us anything very new, but it can be
seen that the ANN is producing a consistent spectral classification
which broadly agrees with a visual analysis of the spectra.
\begin{figure}
\label{fig13}
\psfig{figure=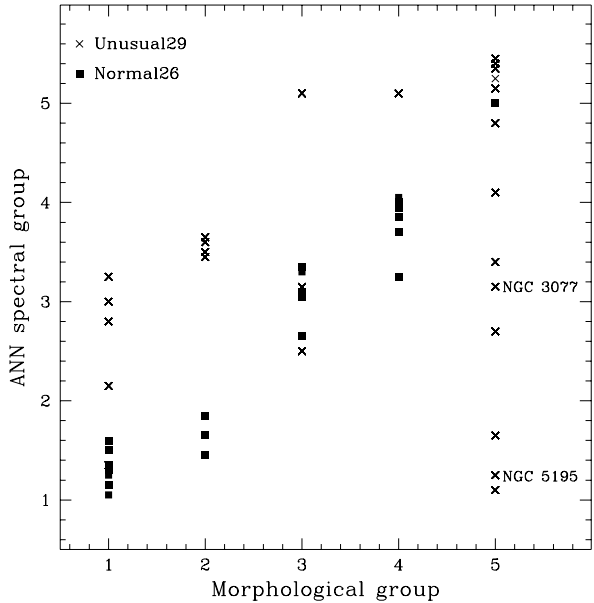,width=3.4in,height=3.4in}
\caption{The Agreement of spectral and morphological classifications
for the Normal26 spectra and the Unusual29 spectra.}
\end{figure}
It can clearly be seen in Figure 13 that the Unusual29 galaxies show
very little agreement between morphological and spectral type, so in
an observed sample, it would be useful to separate the normal from the
unusual spectra. As Figure 7 shows, the Hubble sequence is clearly
evident in relations between the PCs, so it is reasonable to assume
that the unusual spectra do not have this uniformity, for example they
may show discrepant emission and absorption features, or indicate
strong ionization from a Seyfert nuclei without the presence of a
young stellar population. To test this hypothesis, we have taken the
Normal26 and Unusual29 spectra, simulated them at different noise
levels, but without any morphological weighting, and run the PCA
routine on the entire set. We then train the neural net (using 8PCs)
to output 0 if a galaxy is one of the Normal26, or a 1 if the spectrum
is a member of the Unusual29. When the ANN is trained, we ask it to
reclassify the galaxies itself into these two bins. We find that at
$b_{\rm J}=19.7$ about 95\% of the spectra have been classified correctly
in this way.

\section{Discussion}
We demonstrate in this paper that the combination of the PCA technique
and the ANN analysis produces a useful classification tool. The PCA is
useful in three ways in our technique: (i) It allows a transformation
of the data to a more useful set of axes, to reveal segregation in the
sample. (ii) It allows a reconstruction of a low S/N spectrum. (iii) It
provides an economic set of input parameters for the ANN.

We have shown that a limited number of PCs convey the underlying
information in the spectra, and beyond a certain number of PCs
(defined by the noise level) the PCs are not producing useful
information. When the data is restricted to the Normal26 galaxies, the
projections onto the first few PCs reveal segregation in the data
which is chiefly due to the morphological variation in the spectra. If
the Unusual29 spectra are included, the variance in the data is due to
parameters other than purely the morphological type.  We show that the
PCA is best executed on the observed spectra themselves, even if they
are noisy, since the PCs are then highly relevant to that data
set. The alternative approach of projecting the noisy data onto PCs
derived from a small set of high quality spectra can be used, but in
this case, the PCs do not necessarily describe the entire range in the
larger set of observed spectra.

The results from the ANN suggest that a good agreement between
spectral and morphological type can be attained for the Normal26
spectra and that a better classification can be made using this
approach than a simple $\chi^2$ fit to a set of templates. We have
also shown that classification information is not restricted to the
first principal component, such that a classification based purely on
this is not very successful. As expected, little agreement is seen
between the morphology and spectra of the Unusual29 galaxies. One way
to proceed would be to separate these spectra from the main sample,
and then to analyse only the normal galaxies with reference to the
Hubble sequence. We show that it is also possible to use an ANN to
make such a distinction. This would leave a set of unusual spectra
which could be classified separately, or analysed using an alternative
method, such as cluster analysis in the PC space, or one of a variety
of unsupervised data analysis methods which look for trends or
groupings in the data. The ability to highlight unusual spectra would
also prove useful in detecting spectra with bad sky subtraction,
inaccurate fluxing, or incorrect redshift determinations, so that
these could be dealt with separately. The small data set used in this
analysis means we are unable to draw rigorous conclusions as to the
full variation of galaxy spectra. We hope to remedy this situation in
the near future with similar analyses of larger observed data sets
from existing redshift surveys and spectroscopic environmental
studies.

We intend to use the the results of this paper as the basis for
classifying the spectra from the 2dF Galaxy Redshift Survey. We have
shown that a five class classification is obtainable to the proposed
magnitude limit of the survey ($b_{\rm J}=19.7$), but this paper also
demonstrates that it is not necessarily advantageous to restrict the
classification to discovering the morphological types of the
galaxies. It may be better to extend the classification into classes
based entirely on the spectral type. The PCA alone can reveal such
subsets in the data and will provide a powerful tool when used on a
large data set. For the ANN to operate well, a number of the spectra
need to be used as a training set. Several options are available, such
as using morphological classifications from those images bright enough
to be classified by eye, a manual analysis of the high quality spectra
or the use of population synthesis models. A recent investigation
(Sodr\'e \& Cuevas 1996) has successfully related the positions of
observed spectra and model spectra on the PC1/PC2 plane and this
suggests that an ANN trained with model spectra may be able to provide
interesting insights into the physical factors determining galaxy
spectra.

\section{Conclusion}

We have demonstrated a method for the classification of low S/N
spectra using simulations based upon the set of galaxy spectra
presented in Kennicutt (1992). We have developed the simulations to
resemble spectra from the 2dF Galaxy Redshift Survey and show that
reliable classifications, with more than 90\% of the normal galaxies
correctly classified, can be expected to the magnitude limit of the
survey ($b_{\rm J}=19.7$). This may be optimistic, since our small
data set does not encompass the full variation in galaxy spectra, but
our results strongly suggest that the methods in this paper will
provide an interesting analysis technique when the 2dF Galaxy Survey
spectra are available. We have explored the effect of noise on the
Principal Component Analysis of spectra and demonstrate that an ANN is
a useful tool for the classification of noisy spectral data. We show
that the ANN classification is more successful than either a
$\chi^{2}$ template matching approach or a classification based solely
on the projection onto the first principal component. We have also
investigated the agreement of spectral and morphological type and
discussed a method to separate normal from unusual galaxy spectra.

\section*{Acknowledgments}

The authors would like to thank C. Bailer-Jones, R.S. Ellis,
P.J. Francis, J.S. Heyl, M.J. Irwin, A. Naim, L. Sodr\'e, Jr.,
M.C. Storrie-Lombardi, T. von Hippel, and the 2dF Galaxy Survey
collaborators for useful discussions concerning spectral
classification. We would also like to thank B. Ripley for making the
ANN code available.

\appendix
\section[]{Simulations of galaxy spectra}
In order to assess the performance of our classification and
reconstruction procedures in the presence of noise, we have simulated
spectra as they would appear with an 1800 second integration using the
2dF system (Taylor 1994) on the Anglo-Australian telescope.  This
system consists of a prime focus corrector giving a field of view 2
degrees in diameter, within which 400 optical fibres can be accurately
placed at the positions of galaxies.  Each fibre covers a 2'' diameter
region of sky and feeds the light to a highly efficient spectrograph.
Using a 300B grating will give a resolution of 8\AA \ and the present
CCD detectors sample the spectra at 4\AA \ per pixel.  We plan to
survey a large number of galaxies brighter than $b_{\rm J}=19.7$ using this
instrumental set-up, and the simulations are specifically designed to
reproduce the data expected for our survey. However, since the
simulations include all the normal observational effects, they also
provide a reasonable approximation to data for other redshift surveys.

As our starting point we take a sample of high signal-to-noise spectra
(Kennicutt 1992) and redshift each galaxy spectrum to z=0.1, which is
the expected  median redshift for our planned survey.
Then we re-sampled the spectra using 4\AA \ pixels. 
The $b_{\rm J}-V$ colour is calculated from each spectrum so that the $V$
magnitude corresponding to any particular $b_{\rm J}$ can be estimated. 
We scale each spectrum to the relevant $V$ magnitude by first 
normalising it so that the mean flux between 5550\AA \ and 5600\AA \
is unity. 
Then, by considering the flux from a $V=0$ object, the 95100
$cm^{2}$ collecting area of the AAT and an exposure time of 1800
seconds, we calculate the number of galaxy photons at each wavelength. 
Ideally the normalization would be done at 5500\AA \ (being the
characteristic wavelength of the $V$ filter) but at z=0.1 the OIII lines
obscure this area of the continuum and since the continuum is
relatively flat over this short region the difference is minimal.
\begin{figure}
\label{figA1}
\psfig{figure=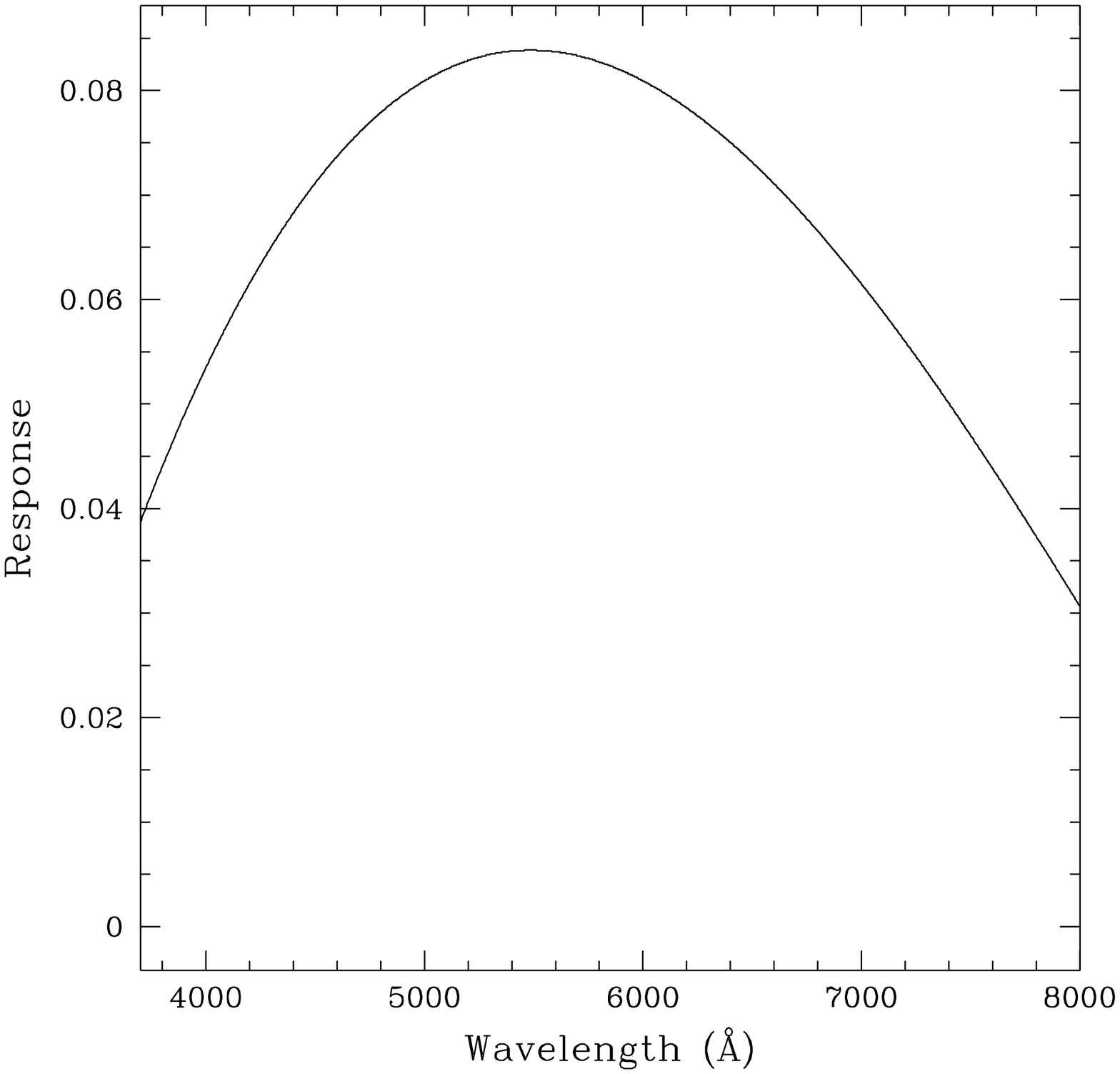,width=3.4in,height=3.0in}
\caption{The response function used for the spectral simulations,
assuming the 300B grating.}
\end{figure}

\begin{figure}
\label{figA2}
\psfig{figure=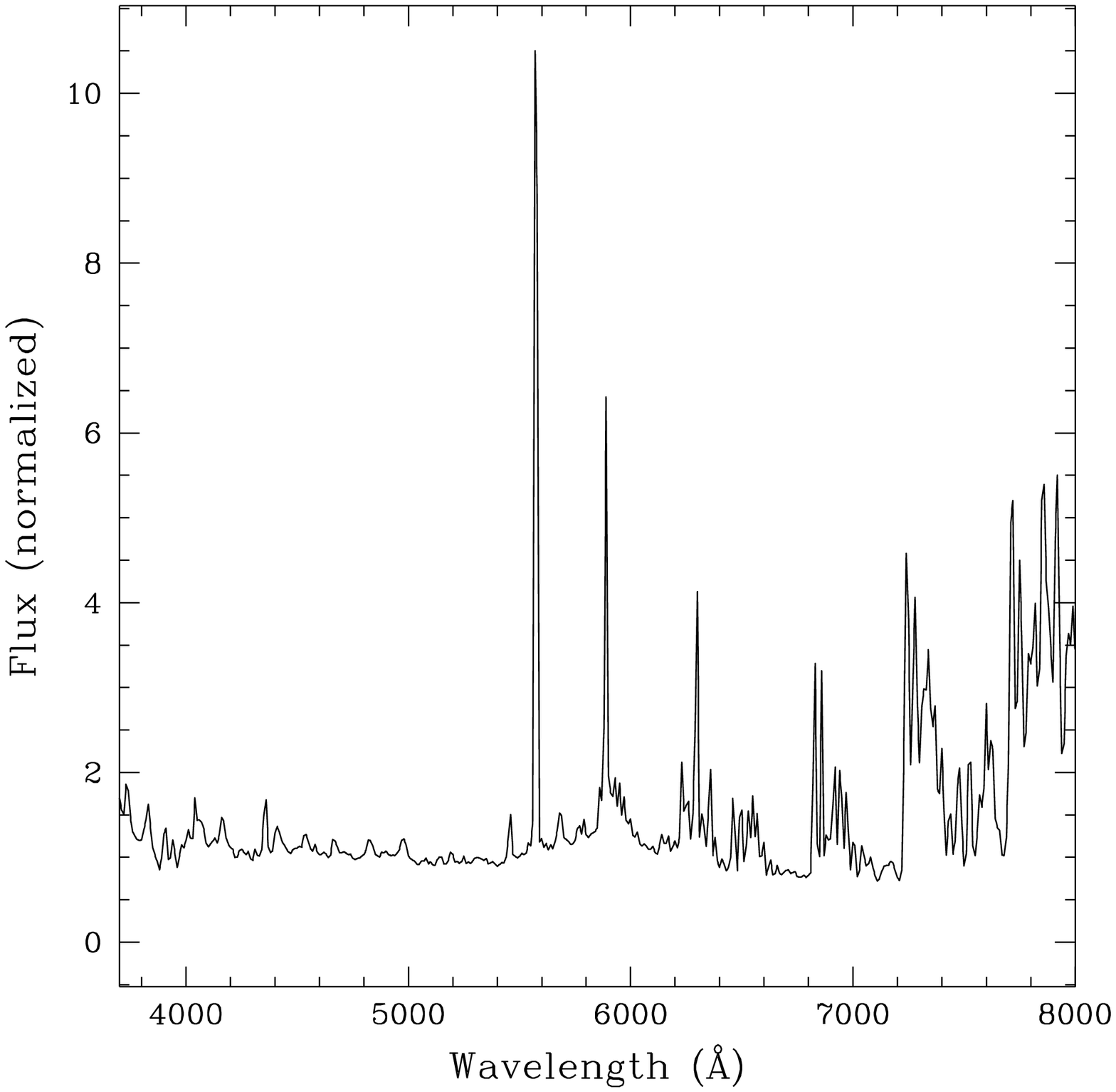,width=3.4in,height=3.0in}
\caption{The sky spectrum used for the spectral simulations.}
\end{figure}
The throughput of the 2dF system has been predicted for a variety of
configurations using detailed modeling of the instrumental optics
(Taylor 1994).  The overall system efficiency predicted for
observations of a point source with zero seeing observed with the 300B
grating is shown as a function of wavelength in Figure A1, and we
multiply each spectrum by this response function.

For galaxies larger than the 2'' diameter fibres, only a fraction of
the total light enters the fibre, and this fraction, $F$, depends on
the surface brightness profile of the galaxy.  We have estimated the
mean value of $F$ as a function of magnitude using the observed
surface brightness profiles of galaxies in the APM galaxy survey
(Maddox et al. 1990). For each galaxy we numerically integrate the
profile out to a radius of 1'', and compare this to the total
magnitude of the galaxy.  At $b_{\rm J}$=19.7 we find that an average
galaxy contains $0.18\pm0.01$ of its total light within a radius of
1'' from the galaxy centre.  Using galaxies with $21<b_{\rm J} <19 $ and
measuring $F$ in 0.2 magnitude bins we find an approximate relation
$$
F=kL^{-0.4} , \eqno (A1)
$$
where $L$ is the galaxy luminosity and $k$ is a constant, set to match
the data at $b_{\rm J}$=19.7. So, for each spectrum we uniformly reduce
the number of photons by the factor $F$ appropriate to each magnitude
 giving our final estimated number of photons per pixel from the galaxy,
$P_{\rm g}$.

A similar procedure is used to calculate the number of photons from
the sky for each pixel. We combined the Kitt peak sky spectrum
(Kennicutt 1992) with a near-IR sky spectrum resulting in a spectrum
with resolution $10-15$\AA \ as shown in figure A2.  We normalised the
spectrum so that the flux at 5500\AA \ is unity, and scaled the
spectrum to give a sky brightness of $B=22.2$ per sq. arc sec., assuming
that $B-V=1.0$ (Smyth 1982). Multiplying by the response as a function
of wavelength and scaling by the area of sky covered by a fibre
finally gives the number of photons received from the sky for each
pixel, $P_{\rm s}$.

\def\var {{\rm var}}

Noise in the spectra is simulated by adding a random deviate to the
predicted number of photons per pixel. Assuming that there are no
systematic errors, the expected variance in the observed number of
photons per pixel is simply given by the Poisson error.  So, for a
pixel with both galaxy and sky, the total number of photons is
$N_{\rm p}=P_{\rm g}+P_{\rm s}$, and the $\var (N_{\rm p}) = P_{\rm g}+P_{\rm s}.$ To estimate the
number of galaxy photons, $N_{\rm g}$, we must subtract an estimate of the
sky photons $N_s$.  If a single sky fibre is used to estimate the sky,
then $ \var (N_{\rm s}) = P_{\rm s} , $ and so the variance in the estimated
number of galaxy photons in a pixel, $N_{\rm g}$, is
$$ \var (N_{ \rm g}) = \var (N_{\rm p} - N_{\rm s}) = P_{\rm g}+2P_{\rm s} . \eqno (A2)$$ 
In practice more than one sky fibre can be used, so the variance in
the subtracted sky spectrum could be reduced to $P_{\rm s}/n$ for $n$ sky
fibres. However, there are other systematic errors, such as variations
in fibre throughput or scattered light, which will not decrease as
more sky fibres are used.  Such effects are impossible to quantify
until full commissioning of the 2dF system takes place, so for this
simulation we have altered equation (A2) to
$$ \var(N_{\rm g}) = P_{\rm g}+1.5P_{\rm s} . \eqno (A3)$$ 
to give an approximation of these factors. 
In practice this uncertainty affects the relation between
magnitude and signal-to-noise by only a few tenths of a magnitude. 
CCD readnoise could be included in these calculations, but for 1800s
exposures, the sky noise vastly dominates the readnoise, so it is
irrelevant for the purpose of this investigation.  So, for each pixel
a random deviation is selected from a Poisson distribution with mean
$P_{\rm g}+1.5P_{\rm s}$, and this deviation is added to the predicted
number of galaxy photons per pixel.  The combination of sky lines and
the response function leads to a large variation in signal to noise as
function of wavelength along the spectrum.

Finally to produce the simulated reduced spectrum, the response
function is divided out, and the spectrum fluxed, de-redshifting to
z=0 (this assumes that the redshift of the spectrum is derived by some
other method), resampled to 4\AA \ bins, and normalized so that the
average flux over the entire wavelength range is unity. Given the
available wavelength range of the Normal26 spectra, this results in
768 bins covering the spectrum between 3712\AA \ and 6780\AA \ , which
includes the strongest optical emission lines.

\section[]{Principal Component Analysis}

A pattern can be thought of as being characterized by a point in an
$M$-dimensional parameter space.  One may wish a more compact data
description, where each pattern is described by $M'$ quantities, with
$ M'\ll M$. This can be accomplished by Principal Component Analysis
(PCA), a well known statistical tool, commonly used in Astronomy
(e.g. Murtagh \& Heck 1987 and references therein).  The PCA method is
also known in the literature as Karhunen-Lo\'eve or Hotelling
transform, and is closely related to the technique of Singular Value
Decomposition.  By identifying the {\it linear} combination of input
parameters with maximum variance, PCA finds $M'$ variables (principal
components) that can be most effectively used to characterize the
inputs.

The first principal component is taken to be along the direction in
the $M$-dimensional input parameter space with the maximum variance.
More generally, the kth component is taken along the maximum
variance direction in the subspace perpendicular to the subspace
spanned by the first $(k-1)$ principal components.

The formulation of standard PCA is as follows.  Consider a set of $N$
objects ($i=1,N$), each with $M$ parameters ($j=1,M$).  If $r_{\rm ij}$ are
the original measurements, we construct normalized properties as
follows:
$$
X_{\rm ij} = { { r_{\rm ij} - {\bar r_{\rm j}} } }, \eqno (B1)
$$
where $ {\bar r_{\rm j}} = { 1\over N} \sum_{\rm i=1}^{\rm N} r_{\rm ij} $ is the mean.
We then construct a covariance matrix
$$
C_{\rm jk} = {1 \over N} \sum_{\rm i=1}^{\rm N} X_{\rm ij} X_{\rm ik} . \hskip1.0cm 1\leq j\leq M \hskip1.0cm 1\leq k\leq M \eqno (B2)
$$
It can be shown that the axis along which the variance is maximal is
the eigenvector ${\bf e_1}$ of the matrix equation
$$
C {\bf e_1} = \lambda_1 {\bf e_1}, \eqno (B3)
$$
where the $\lambda_1$ is the largest eigenvalue, which is in fact the
variance along the new axis.  The other principal axes and
eigenvectors obey similar equations.  It is convenient to sort them
in decreasing order, and to quantify the fractional variance by
$\lambda_{\rm \alpha} / \sum_{\rm \alpha} \lambda_{\rm \alpha}$.  It is also
convenient to re-normalize each component by $\sqrt \lambda_{\rm \alpha}$,
to give unit-variance along the new axis.  We note that the weakness
of PCA is that it assumes linearity and also depends on the way the
variables are scaled. The matrix of all the eigenvectors forms a new
set of orthogonal axes which are ideally suited to a description of
the data set. A truncated eigenvector matrix using only `P'
eigenvectors

$$
U_{\rm P}=\{e_{\rm jk}\} \hskip2.0cm 1\leq k\leq P \hskip1.0cm 1\leq
j\leq M \eqno (B4)
$$
can be constructed where $e_{\rm jk}$ is the jth component of the kth
eigenvector.

Now if we take a specific spectrum from the matrix defined in equation
(B1), or possibly a spectrum from a different source which has
been similarly normalized and mean subtracted, it can be
represented by the vector of fluxes ${\bf x}$. We can find the
projection vector ${\bf z}$ onto the M principal components from

$$
{\bf z}={\bf x}U_{\rm M} . \eqno (B5)
$$
Multiplying by the inverse, the spectrum is given by
$$
{\bf x}={\bf z}{U_{\rm M}}^{\rm -1}={\bf z}{U_{\rm M}}^{\rm t} , \eqno (B6)
$$ 
since $U_{\rm M}$ is an orthogonal matrix by definition. However, if
we were to use only P principal components the reconstructed
spectrum would be
$$
{\bf x}_{rec}={\bf z}{U_{\rm P}}^{\rm t} , \eqno (B7)
$$ 
which is an approximation of the true spectrum.

\section[]{Artificial Neural Networks}
It is common in Astronomy and other fields to fit a model with several
free parameters to observations. This regression is usually done by
means of $\chi^2$ minimization. A simple example of a model is a
polynomial with the coefficients as the free parameters. Consider now
the specific problem of morphological classification of galaxies. If
the type of the $i$th galaxy is $T_i$ (e.g. on the numerical system
$[-5,10]$), and we have a set of parameters ${\bf x}_i$ (e.g. isophotal
diameters, colours or spectral lines) then we would like to find the
free parameters ${\bf w}$ (`weights') such that the cost function
$$
E = {1 \over 2} \sum_{i=1}^N [T_i - F({\bf w}, {\bf x}_i) ] ^2, \eqno (C1)
$$
is minimized. The non-linear function $F({\bf w}, {\bf x}) $
represents the `network', which consists of a set of input
nodes, a set of output nodes and one or more further layers of
'hidden' nodes between the input and output layers. The
`hidden-layers' allow curved boundaries around clouds of data points
in the parameter space.

For a given network architecture the first step is the `training' of
the ANN.  In this step the weights $w_{ij}$'s (the `free parameters')
are determined by minimizing `least-squares'.  Each node (except the
input nodes) receives the output of all nodes in the previous layer
and produces its own output, which then feeds the nodes in the next
layer.  A node at layer $s$ calculates a linear combination over the n
inputs $x_i^{(s-1)}$ from the previous layer $s-1$ according to
$$
I_j^{(s)} = \sum_{i=0}^n\; w_{ij}^{(s)} \; x_i^{(s-1)} \eqno (C2)
$$
where the $w_{ij}$'s are the weights associated with that node.
Commonly one takes $x_0=1$, with $w_{0j}$ playing the role of a `bias'
or DC level.  The node then fires a signal
$$
x_j^{(s)} = f(z), \eqno (C3)
$$ 
where $z$ here stands for $I_j^{(s)}$, and $f$ is a non-linear
transfer function usually of the sigmoid form
$$
f(z) = 1 /[1+ \exp(-z)] \eqno (C4)
$$ 
in the interval [0,1], or
$$
f(z) = \tanh (z) \eqno (C5)
$$ 
in the interval [-1,1].

For each galaxy in the training set, the network compares the output F
with the desired type T to give a cost function (equation C1) averaged
over all the training galaxies. This is minimized with respect to free
parameters, the weights $w_{ij}$.

The interpretation of the network output depends on the network
configuration.  For example, a single output node provides a
continuous output while several output nodes can be used to assign
probabilities to different classes (e.g.  5 morphological types of
galaxies, e.g. Storrie-Lombardi et al. 1992; Lahav et al. 1996).  There
are various optimization algorithms for finding the weights.  One
popular algorithm is the Backpropagation (e.g. Rumelhart, Hinton \&
Williams 1986; Hertz et al. 1991), where the minimization is done by
the chain rule (gradient descent).

A more efficient method, used in our study, is Quasi-Newton.  In
short, the cost function $E({\bf w})$ in terms of the weights vector
${\bf w}$ is expanded about a current value ${\bf w_0}$:

$$
E({\bf w}) = E({\bf w_0}) + ({\bf w} - {\bf w_0}) \; \nabla E({\bf
w_0}) \; + {1 \over 2} ({\bf w} - {\bf w_0})\cdot {\bf H} \cdot ({\bf
w} - {\bf w_0}) + ..., \eqno (C6)
$$
where ${\bf H}$ is the Hessian with elements $H_{ij} = {\partial^2 E
 \over { \partial w_i \partial w_j}}$ evaluated at ${\bf w_0}$.  The
 minimum approximately occurs at
$$
\nabla E ({\bf w}) \approx \nabla E ({\bf w_0}) + {\bf H} \cdot ({\bf
w} - {\bf w_0}) = 0.  \eqno (C7)
$$  
Hence an estimation for the optimal weights vector is at
$$
{\bf w} = {\bf w_0} - {\bf H}^{-1} \nabla E ({\bf w_0}).  \eqno (C8)
$$ 
In the standard Newton's method a previous estimate of ${\bf w}$ is
used as the new ${\bf w_0}$.  Calculating the Hessian exactly is
expensive computationally, and in the quasi-Newton method an iterative
approximation is used for the inverse of the Hessian (e.g. Press et
al. 1992; Hertz et al. 1991).

The determination of many free parameters, the weights $w_i$'s in our
case, might be unstable.  It is therefore convenient to regularise the
weights, e.g. by preventing them from growing too much. In the ANN
literature this is called `weight decay'.  This approach is analogous
to Maximum Entropy, and can be justified by Bayesian arguments, with
the regularising function acting as the prior in the weight space.
One possibility is to add a quadratic prior to the cost function and
to minimize
$$
E_{tot} = \alpha E_w + \beta E_D, \eqno (C9)
$$
where $E_D$ is our usual cost function, based on the data
(e.g. equation C1) and
$$
E_w = {1 \over 2} \sum_{i=1}^Q w_i^2 \eqno (C10)
$$
is the chosen regularising function, where $Q$ is the total number of
weights.  The coefficients $\alpha$ and $\beta$ can be viewed as
`Lagrange multipliers'.  While sometime they are specified ad-hoc, it
is possible to evaluate them `objectively' by Bayesian arguments in
the weight-space.  Approximately $\alpha^{-1} \approx 2 {\hat E_w}/Q$
and $\beta^{-1} \approx 2 {\hat E_D} /N$, (e.g. MacKay 1992, Ripley
1993, Lahav et al. 1996).  We note that this analysis makes sense if
the input and output are properly scaled e.g. between [0, 1] with
sigmoid transfer functions, so all the weights are treated in the
regularisation process on `equal footing'.  It can be generalized for
several regularising functions, e.g. one per layer.

We note that the addition of the regularisation term $E_w$ changes the
location of the minimum, now satisfying
$$
\nabla E_D = - { \alpha \over \beta} \nabla E_w = - { \alpha \over
\beta} {\bf w}, \eqno (C11)
$$ 
as from equation (C10) $\nabla E_w = {\bf w}$. The effect of the
regularisation term here is reminiscent of the restoring force of an
harmonic oscillator: the larger ${\bf w}$ is the more it will get
suppressed.  We applied least-square minimization using a Quasi-Newton
method as implemented in a code kindly provided to us by B.D. Ripley.

\bsp

\end{document}